\documentclass{aastex}
\usepackage{emulateapj5}

\shorttitle{Fragmentation of Magnetically Subcritical Clouds}
\shortauthors{Li \& Nakamura}

\begin{document}

\title{Fragmentation of Magnetically Subcritical Clouds into Multiple
Supercritical Cores and the Formation of Small Stellar Groups}

\author{Zhi-Yun Li}
\affil{Department of Astronomy, University of Virginia, P.O. Box 3818,
Charlottesville, VA 22903; zl4h@virginia.edu}
\and
\author{Fumitaka Nakamura}
\affil{Faculty of Education and Human Sciences, Niigata University,
8050 Ikarashi-2, Niigata 950-2181, Japan, and Astronomy Department,
University of California at Berkeley, Berkeley, CA 94720}

\begin{abstract}

Isolated low-mass stars are formed in dense cores of molecular clouds. 
In the standard picture, the cores are envisioned to condense out of 
strongly magnetized clouds through ambipolar diffusion. Most previous 
calculations based on this scenario are limited to axisymmetric cloud 
evolution leading to a single core, which collapses to form an isolated 
star or stellar system at the center. These calculations are here 
extended to the nonaxisymmetric case under thin-disk approximation, 
which allows for a detailed investigation into the process of fragmentation, 
fundamental to binary, multiple system, and cluster formation. We have 
shown previously that initially axisymmetric, magnetically subcritical 
clouds with an $m=2$ density perturbation of modest fractional amplitude 
($\sim 5\%$) can develop highly elongated bars, which facilitate binary 
and multiple system formation. In this paper, we show that in the presence 
of higher 
order ($m\ge 3$) perturbations of similar amplitude such clouds are 
capable of breaking up into a set of discrete dense cores. These multiple
cores are magnetically supercritical. They are expected to collapse into 
single stars or stellar systems individually and, collectively, to form 
a small stellar group. Our calculations demonstrate that the standard 
scenario for single star formation involving magnetically subcritical 
clouds and ambipolar diffusion can readily produce more than one star, 
provided that the cloud mass is well above the Jeans limit and relatively 
uniformly distributed. The fragments develop in the central part 
of the cloud, after the region has become magnetically supercritical 
but before rapid collapse sets in. It is enhanced by the flattening
of mass distribution along the field lines and by the magnetic tension 
force, which is strong enough during the subcritical-to-supercritical 
transition to balance out the gravity to a large extent and thus 
lengthen the time for perturbations to grow and fragments to separate
out from the background. 
\end{abstract}


\section{Introduction}
\label{sec:introduction}

Dense cores of molecular clouds play a pivotal role in star formation
(Myers 1999). They provide a crucial link between the molecular clouds
and the stars formed in them. In the standard picture for isolated
low-mass star formation, the cores are envisioned to gradually condense
out of a magnetically subcritical background cloud, through ambipolar
diffusion (Shu, Adams \& Lizano 1987; Mouschovias \& Ciolek 1999; see
Nakano 1998 and Myers 1999 for an alternative view involving turbulence
decay). Detailed calculations based on this scenario have been carried
out by many authors (e.g., Nakano 1979; Lizano \& Shu 1989; Ciolek \& 
Mouschovias 1993; Basu \& Mouschovias 1994). It has been established
that prior to star formation dense cores have (1) a central region of
flat density distribution surrounded by a roughly $r^{-2}$ envelope,
(2) an infall speed over an extended region that could be a significant
fraction of the isothermal sound speed, and (3) a field strength
typically half the critical value. All these features are consistent
with the observations of L1544 (Tafalla et al. 1998; Ward-Thompson et 
al. 1999; Crutcher \& Troland 2000), arguably the best studied starless
core (e.g., Caselli et al. 2002). Ambipolar diffusion-driven cloud
evolution models computed specifically for this source (without explicit
treatment of turbulence) can match quantitatively its observed mass
distribution, velocity field and magnetic field strength (Ciolek \& Basu
2000), and plausibly the abundances and spatial distributions of various
commonly-observed molecular species, such as CO, N$_2$H$^+$, and CCS,
as well (Li et al. 2002). For L1544, the standard scenario appears
to provide a reasonable description (see Crutcher et al. 1994 for
the successful application of a similar model to the dark cloud B1, for
which the field strength is also measured).

Most calculations based on the standard scenario are axisymmetric. As
such, they can not directly address the fundamental issue of cloud
fragmentation, which lies at the heart of binary, multiple stellar
system and cluster formation (Bodenheimer et al. 2000). Indeed, there
has been some nagging concern whether a magnetic field strong enough 
to provide most of the cloud support would at the same time prevent 
fragmentation (Galli et al. 2001). This is certainly the case if the 
field is completely frozen in the matter (Nakano 1988). Even a somewhat
weaker frozen-in field which does not prevent a cloud from collapsing
dynamically may stifle fragmentation (Dorfi 1982; Phillips 1986a,b).
However, these results have limited applicability to
molecular clouds, which are only lightly ionized and partially coupled
to the magnetic field. For a lightly ionized medium of uniform density
threaded by a uniform magnetic field, Langer (1978) showed through
linear analysis that the criterion for gravitational instability is
unaffected by the field, although the growth rate can be. For a strongly
magnetized dark cloud, the instability grows on an ambipolar diffusion
time scale, which is typically an order of magnitude longer than the
dynamic time. Nonlinear developments of this magnetically-mediated
gravitational instability in molecular clouds have not been explored
in any detail (see Indebetouw \& Zweibel 2000 for simulations of a
related instability). How or even whether they can lead to cloud 
fragmentation into discrete pieces capable of forming more than one 
star remains uncertain. It is the focus of our investigation.

There is some indication that a strong magnetic field may actually promote
molecular cloud fragmentation. Boss (2002) followed the evolution of a
set of initially magnetically supported clouds in three dimensions (3D),
taking into account several magnetic effects approximately. He concluded
that the cloud fragmentation is enhanced by magnetic fields because the
magnetic tension helps to
prevent a central density singularity from forming and producing a
dominant single object. Li (2001) studied the 1D evolution of a set
of flattened, magnetically subcritical clouds assuming axisymmetry
and found that either a dense supercritical core or off-centered 
ring forms as a result of ambipolar diffusion. Nakamura \& Li (2002) 
showed through
2D nonaxisymmetric calculations that the core-forming clouds are
unstable to the $m=2$ nonaxisymmetric mode, with density perturbations 
of modest fractional amplitude ($\sim 5\%$) growing nonlinearly into 
bars of a typical aspect ratio $\sim 2$ during the transition period 
after the core has just become supercritical (which makes the bar 
growth possible) but before rapid collapse sets in (which leaves little 
time for further growth). They found that, by the time the isothermal
approximation starts to break down, the elongation has been strongly 
amplified by the Lin-Mestel-Shu (1965) instability, producing highly 
elongated bars. These bars are expected to break
up gravitationally into pieces during the subsequent adiabatic phase of
evolution, which could lead to the formation of binary and multiple 
stellar systems. The possible bar fragmentation into multiple
systems will be explored elsewhere. Here, we concentrate on the
evolution of nonaxisymmetrically perturbed, ring-forming clouds which,
as we show in the paper, are capable of producing a number of discrete,
magnetically supercritical cores while still in the isothermal phase
of cloud evolution. Even though we cannot follow the evolution of the
cores beyond the isothermal phase, we anticipate each of them to collapse 
individually into a single star or stellar system and, collectively, 
to form a stellar group or small cluster. 

We describe our formulation of the problem of magnetic cloud evolution 
and fragmentation, including the governing 
equations, initial conditions, and numerical methods in \S~2. This 
is followed by a set of representative models illustrating the main
features of nonaxisymmetric cloud evolution leading to fragmentation
(\S~3). In the last section (\S~4), we discuss the nature of magnetic 
cloud fragmentation, comment on the implications of our calculations 
on stellar group formation, and conclude.

\section{Formulation of The Problem}
\label{sec:method}

\subsection{Basic Equations}
\label{equations}

We consider nonaxisymmetric evolution of strongly magnetized molecular
clouds driven by ambipolar diffusion, as envisioned in the standard
picture of low-mass star formation. The strong magnetic field allows
the cloud material to settle along field lines into a disk-like geometry,
maintaining vertical force balance even during the dynamic collapse
phase of cloud evolution (Fiedler \& Mouschovias 1993). We adopt the
standard thin-disk approximation
\citep[e.g.,][]{GCiolek93, SBasu94, FShu97, FNakamura97, RStehle01, ZLi01},
and follow the 2D cloud evolution in the $x$-$y$ plane in a Cartesian
coordinate system $(x,y,z)$. Our 2D model is the nonaxisymmetric extension
of the 1D model of \citet{ZLi01}. We assume that the disk
is threaded by an ordered magnetic field which is current-free
outside the disk and which becomes uniform far from the cloud with a
constant strength $B_{z, \rm \infty}$. Twisting of field lines is 
possible, as a
result of, e.g., magnetic braking; it is ignored here since the type
of fragmentation that we will discuss is gravitational in origin and
does not rely on rotation 
(see \S~\ref{norotation}). Magnetic braking of
the disk and the associated field twisting can be included in our
formulation approximately if so desired (Basu \& Mouschovias 1994).

The differential equations governing the evolution of a strongly magnetized
disk are given in a vertically integrated form. Mass conservation of the
disk is expressed by
\begin{equation}
\frac{\partial \Sigma }{\partial t} +
\nabla \cdot \left(\Sigma \mbox{\boldmath$V$} \right) = 0 \; ,
\label{eq:basic1}
\end{equation}
where $\Sigma$, $t$, $\mbox{\boldmath$V$} = \left(V_x, V_y \right)$
are, respectively, the surface density, time, and velocity in the disk.
The momentum equation is given by
\begin{equation}
 \frac{\partial \left(\Sigma \mbox{\boldmath$V$}\right)}{\partial t} +
  \nabla \left(\Sigma \mbox{\boldmath$V$}\mbox{\boldmath$V$} \right)
   + \nabla P + H \nabla \left(\frac{B_z ^2}{4\pi}\right)
   - \frac{B_z \mbox{\boldmath$B$}}{2\pi} + \Sigma \mbox{\boldmath$g$}= 0 \; ,
 \label{eq:basic2}
\end{equation}
where $\mbox{\boldmath$B$} = \left(B_x, B_y \right)$,
$\mbox{\boldmath$g$} = \left(g_x, g_y \right)$, and $P$ is the
vertically integrated pressure.
The equation for the evolution of the vertical field component is
\begin{equation}
 \frac{\partial B_z}{\partial t} +
  \nabla \cdot \left(B_z \mbox{\boldmath$V_{\rm B}$} \right) = 0 \; ,
 \label{eq:basic3}
\end{equation}
where $\mbox{\boldmath$V_{\rm B}$}=(V_{{\rm B},x},V_{{\rm B},y})$ is the
velocity vector of magnetic field lines in the disk plane.

The governing equations (\ref{eq:basic1})-(\ref{eq:basic3}) are 
supplemented by the vertically integrated equation of state
\begin{equation}
P = c_s^2 \Sigma ,
\label{eq:EOS}
\end{equation}
where $c_s$ is the (effective) isothermal sound speed, which we take
to be a constant since we are interested in fragmentation during the
isothermal regime; the transition to optically thick, adiabatic regime 
of cloud
evolution will be treated elsewhere. We do not consider explicitly
in the equation of state the presence and possible decay of turbulence,
although some of its effects can be incorporated into the effective
sound speed $c_s$ (Lizano \& Shu 1989). The ``turbulent'' pressure
should have little direct effect on the process
of fragmentation, which occurs at relatively high densities where the
turbulent motions are observed to be subsonic (Myers 1999). It does have
a great influence on the initial conditions for our cloud evolution,
given in the present study by prescription (see \S~\ref{condition}).

In addition, we can relate the disk half-thickness to the mass density
$\rho$ through
\begin{equation}
H = \Sigma /(2\rho).
\label{eq:thickness}
\end{equation}
Assuming hydrostatic equilibrium in the vertical direction of
the disk, we find
\begin{equation}
\rho = \frac{\pi G\Sigma ^2}{2 c_s^2}\left(1+\frac{B_x^2 + B_y^2}{4\pi
^2 G \Sigma^2}\right) + \frac{P_e}{c_s^2} ,
 \label{eq:volumndensity}
\end{equation}
to the lowest order in $(H/r)$, where the cylindrical radius
$r=(x^2+y^2)^{1/2}$.
The two terms in the brackets represent, respectively, the gravitational
compression and magnetic squeezing of the disk material.
The quantity $P_e$ is the ambient pressure
 that helps confine the disk, especially in
low column density regions where gravitational compression
 is relatively weak.

In a lightly ionized medium such as molecular cloud,
the field lines slip relative to the neutral matter at a velocity
\begin{equation}
  \mbox{\boldmath$V_{\rm B}$} - \mbox{\boldmath$V$}
   = {t_c \over \Sigma}  \left[\frac{B_z \mbox{\boldmath$B$}}
	{2\pi}- H \nabla \left(\frac{B_z^2}{4\pi}\right) \right] ,
 \label{eq:basic4}
\end{equation}
where the coupling time between the magnetic field and
neutral matter, $t_c$, is given approximately by
\begin{equation}
t_c = \frac{1.4}{\gamma C\rho^{1/2}} \; ,
 \label{eq:basic5}
\end{equation}
in the simplest case where the coupling is provided by ions that are well 
tied to the field lines and the ion density $\rho _i$ is related to the
cloud density $\rho$ by the canonical expression
$\rho _i = C\rho ^{1/2}$. Here, $\gamma C = 1.05\times 10^{-2}$
cm$^{3/2}$g$^{-1/2}$s$^{-1}$ \citep{FShu91}
and the factor 1.4 comes from the fact
that the cross section for ion-helium collision is small compared to
that of ion-hydrogen collision \citep{TMouschovias91}. More detailed
treatments of the magnetic coupling including the effects of dust
grains are possible (e.g., Nishi, Nakano \& Umebayashi 1991). We
postpone such treatments to a future study.

The gravitational acceleration $\mbox{\boldmath$g$}$ and field strength
$\mbox{\boldmath$B$}$ in the disk plane are determined from the
gravitational potential $\psi_G(x,y,z)$ and magnetic potential $\psi
_B(x,y,z)$, which satisfy
\begin{equation}
\nabla ^2 \psi_G = 4\pi G\Sigma \delta (z) ,
\end{equation}
and
\begin{equation}
\nabla ^2 \psi _B = (B_z - B_{z,\infty})\delta (z) ,
\label{eq:basic6}
\end{equation}
where $\delta$ is the Dirac delta function.

\subsection{Initial Conditions}
\label{condition}

The initial conditions for the core formation are not well determined
either observationally or theoretically. Following \citet{SBasu94} and
\citet{ZLi01}, we prescribe a reference state in which the clouds are
axisymmetric, slowly rotating, and 
are threaded by a uniform magnetic field, $B_{z,
\rm ref} (x,y) = B_{z, \infty}$.  The reference surface density
and rotation velocity distributions are taken, respectively, to have 
the form 
\begin{equation}
\Sigma _{\rm ref} (x,y) = \frac{\Sigma_{\rm 0,ref}}
{\left[1+(r/r_0)^n\right]^{4/n}} ,
\label{eq:basic11}
\end{equation}
and
\begin{equation}
V_{\phi, \rm ref} (x,y)={ 4 \ \omega \ r \over r_0
	+\sqrt{r_0^2+r^2} } c_{s,\rm{eff}}  ,
\label{eq:basic12}
\end{equation}
where the exponent $n$ controls the surface density profile and (indirectly) 
the amount of mass in the central plateau region where the distribution is
more or less uniform, and the dimensionless
parameter $\omega$ measures the rotation rate relative to $c_{s,\rm{eff}}
=c_s (1+\Gamma_0^2)^{1/2}$, which is essentially the magnetosonic speed 
(the magnetic contribution comes in through the parameter $\Gamma_0$ to
be defined in the next paragraph). 
The somewhat arbitrary profile (\ref{eq:basic11}) 
is chosen mainly to minimize numerical edge effects; 
it should not be confused with the
plateau-plus-envelope density distribution observed in several starless
dense cores (Ward-Thompson et al. 1999; Shirley et al. 2000) and used
as the initial conditions for some cloud collapse calculations (e.g.,
Boss 2002)---the observed profile develops naturally in the central
part of our model cloud later on through ambipolar diffusion-driven
evolution. 
Note that the rotational velocity prescribed by equation (\ref{eq:basic12}) 
increases linearly with radius inside $r_0$ before asymptoting to a 
constant value at larger radii.

The background field strength is characterized by the dimensionless
flux-to-mass ratio $\Gamma_0=B_{z, \infty}/(2\pi G^{1/2} 
\Sigma_{\rm 0,ref})$,
which must be greater than unity for the cloud to be magnetically
subcritical. The reference clouds are not in an equilibrium state and
are allowed to evolve into one with magnetic field frozen-in, before
ambipolar diffusion is turned on at time $t=0$. To save computational 
time, we obtain the initial, axisymmetric equilibrium state using the 
1D code of \citet{ZLi01}, in which the cloud radius is chosen to be 
twice the characteristic radius $r_0$ for numerical convenience. At 
$t=0$, we superpose on top of the equilibrium column density $\Sigma_0(x,y)$ 
a nonaxisymmetric perturbation of fractional 
amplitude $A$, so that
\begin{equation}
\Sigma (x,y)= \Sigma_0(x,y)\ [1 + A\ \cos(m \phi)],
\end{equation}
where $\phi$ is the azimuthal angle measured from the $x$-axis. Both
the m=2 bar mode and higher order modes will be considered in the
present study, as well as a form of random perturbation (see 
\S \ref{random}). 
The subsequent, ambipolar diffusion-driven evolution of the perturbed
cloud is followed numerically, subject to the condition of fixed
external pressure and $B_z$ at the cloud outer boundary. 

\subsection{Numerical Methods}
\label{methods}

Our calculations are carried out using an MHD code based on that of
\citet{FNakamura97}. The hydrodynamic part is solved using Roe's
TVD (Total Variation Diminishing) method given in \citet{CHirsch90}.
To achieve the second-order accuracy, we implemented the so-called 
``MUSCL'' (Monotone Upstream-centered Schemes for Conservation Laws) 
approach, in which the physical quantities between two adjacent cells 
are extrapolated to the cell boundary and the linearized Riemann 
problem is solved exactly for the numerical fluxes at the cell boundaries.
The minmod limiting function is used to satisfy the TVD condition.
The magnetic part is treated as follows. The equation for magnetic
field evolution has a form identical to that for mass conservation,
and is solved in the same manner.
Both the magnetic and gravitational potentials satisfy
the Poisson equation with a source term containing
$\delta (z)$, and both of them are solved using a
convolution method described in \citet{RHockney81}.
The time step is determined by the Courant-Friedrich-Levy (CFL) condition,
with a contribution from the magnetic field included \citep[see also]
[]{RStehle01}. The original code has been tested with many test problems, 
including 1D shock tube and point explosion. The 
tests have verified that the code has
second-order accuracy and produces no oscillations of numerical origin
\citep[e.g.,][]{TMatsumoto97}. In addition, we have followed the evolution 
of a set of unperturbed axisymmetric clouds with both the 2D code 
developed here and the 1D code of \citet{ZLi01}. The cloud evolution
time, surface density distribution, magnetic field strength and 
gravitational acceleration obtained with the two codes all agree to 
within a few percent.

Following \citet{FNakamura99}, we employ a zooming technique to resolve 
the dense, small-scale structures that develop during the cloud evolution. 
We refine and zoom-in the computational grid whenever the so-called
Jeans condition is about to be violated \citep{KTruelove97}, taking 
into account the magnetic effect on Jeans condition. For production
runs, we start all simulations on a $256\times 256$ uniform, rectangular 
grid. The grid points are doubled in each direction at the first 
refinement, without changing the size of the simulation box. At each 
subsequent refinement, the computational domain for hydrodynamics 
is shrunk by half in each direction, leaving the grid number unchanged 
($512\times 512$). All physical quantities at the domain boundary 
are fixed during a given level of (subsequent) refinement.
In our calculations, this boundary condition is identical to an inflow
boundary condition because, when refinements are needed, the cloud 
mainly collapses toward the central high density region (although some 
transient outflows are possible). All physical quantities on the refined 
grids are obtained by linear interpolation. As a convergence test, we 
have doubled the initial grid number from $256\times 256$ to $512\times 
512$ for a ``standard'' model (see Table~1 and \S~\ref{rotation}) and 
found no significant differences. In a future study, we plan to take a 
nested-grid approach, rather than zooming, which should improve the 
boundary condition. 

The actual numerical computations are carried out using non-dimensional
quantities. The units we adopted are $c_s=1.88\times 10^4 T_{10}^{1/2}
$~cm~s$^{-1}$ for speed (where $T_{10}$ is the effective temperature in
units of 10~K), $\Sigma _{0,\rm ref}=4.68\times 10^{-3} A_V$~g~cm$^{-2}$
for surface density (where $A_V$ is the visual extinction vertically 
through the disk for standard grain properties), and $B_{z,\infty}
=7.59\; A_V\; \Gamma_0$~$\mu$G for magnetic field strength (where 
$\Gamma_0$
is the flux-to-mass ratio in units of the critical value). The units for
length, time, and mass are, respectively, $L_0\equiv c_s^2/(2\pi G
\Sigma_{0,\rm ref})=5.83\times 10^{-2} T_{10}/A_V$~pc, $t_0\equiv c_s
/(2\pi G\Sigma_{0,\rm ref})=3.03\times 10^5 T_{10}^{1/2}/A_V$~yr, and
$M_0\equiv \Sigma_{0,\rm ref}L_0^2=7.64\times 10^{-2} T_{10}^2/A_V
$~$M_\odot$. We normalize the external pressure by 
$2\pi G \Sigma_{0,\rm ref}^2$,
and set its dimensionless value to 0.025 for all models.

\section{Numerical Results}
\label{results}

The number of parameters needed to completely specify the model cloud and
imposed perturbation is rather large. These include the characteristic
radius $r_0$, exponent $n$ in the reference surface density distribution,
rotation parameter $\omega$, and dimensionless flux-to-mass ratio
$\Gamma_0$ for the cloud, and the order $m$ and fractional amplitude $A$
for the perturbation. We have carried out a systematic survey of these
parameters as well as a form of random perturbation (of mixed modes), 
and selected the models listed in Table~1 to illustrate the salient 
features of magnetic cloud fragmentation. 

\subsection{Cloud Fragmentation Without Rotation}
\label{norotation}

To begin with, we consider a non-rotating ($\omega=0$) cloud with the
reference state specified by a set of parameters $r_0=5\pi$, $n=8$,
and $\Gamma_0=1.5$. The cloud is therefore magnetically subcritical,
with the field strength $50\%$ above the critical value at the center.
This set of parameters is chosen because, in the absence of any
nonaxisymmetric perturbations, the cloud so specified would collapse to
produce a dense supercritical ring, which is prone to fragmentation.
The reference cloud is allowed to evolve into an equilibrium
configuration, with the magnetic
field frozen-in. After the equilibrium is reached, we reset the time
to $t=0$, and apply an $m=4$ perturbation to the surface density
distribution, with a fractional amplitude of $A=0.05$. The evolution of
such nonaxisymmetrically perturbed cloud is followed numerically to
progressively higher densities and smaller scales. The results are
shown in panels (a) though (e) of Fig.~1 and described as follows.

As is usual with ambipolar diffusion-driven evolution, the cloud spends
most of its time in the subcritical phase when the dimensionless
minimum flux-to-mass ratio (at the density peak) $\Gamma > 1$ (see 
panels a and b of Fig.~1).
During this period, inward and outward motions coexist. Both are
very subsonic, and exhibit an overall oscillatory pattern (of small
amplitude). The oscillation signifies that the cloud is stable to the
added $m=4$ perturbation during the subcritical phase, in agreement with
linear analysis by Nakano (1988) and others. Indeed, we have verified
numerically that for a frozen-in field the subcritical cloud remains
stable to this and other modes. Once the
minimum flux-to-mass ratio has dropped below the critical value, the
cloud evolution enters a qualitatively new phase. Its contraction
accelerates, with speeds first approaching, and eventually exceeding,
the sound speed (see panels c-e). It is during this phase that
fragmentation occurs. The beginning stage of the fragmentation is
shown in panel c), where the velocity vectors deviate significantly
from the radial direction in some regions, creating appreciable 
distortions in the surface density distribution. By the time shown in
panel d), the inflow appears to have converged preferentially toward
individual pockets, creating four well-defined blobs or fragments.
These blobs grow quickly in (surface) density, and become cleanly 
separated
from the background---a common envelope---and from one another by the
end of the computation (panel e), when the peak surface density has
increased over its reference value by more than a factor of 600.
The high density blobs resemble in every way the single (starless)
cores obtained in previous calculations, except of course the
presence of neighbors; they are simply multiple dense 
cores\footnote{Operationally, we identify any local maximum in 
surface density enclosed within one or more closed contours as
a ``core'', even though the core may be highly elongated (and
thus more filament-like) and, in the case of modest contrast with 
background, its existence depends on the number of contour levels chosen.}.

The multiple cores shown in panel e) are formed together, at separations
nearly two orders of magnitude smaller than the central plateau part of
the original
cloud, from which the cores are condensed out of. The cores are clearly
supercritical, with a minimum flux-to-mass ratio of only $\Gamma=0.62$.
They are collapsing dynamically, and are well on their way to forming
individual stars or stellar systems. Note that the cores are significantly
elongated, with an aspect ratio of roughly 3. The degree of elongation
should increase rapidly before density singularities develop due to
Lin-Mestel-Shu (1965) instability, forming highly
elongated bars by the time the isothermal approximation starts to break
down, as in the collapse of single, isolated cores studied by Nakamura
\& Li (2002). We expect the bars to break up during the subsequent
adiabatic phase of cloud evolution, each forming perhaps a binary or
multiple system. Together, we anticipate the production of a small
stellar group in the cloud as a likely outcome, although we are unable
to follow the evolution further because of the increasing demand on
spatial resolution.
%
%

The properties of cloud fragmentation 
depend somewhat on the flux-to-mass ratio of the reference state
$\Gamma_0$ ($ > 1$). Exact values of $\Gamma_0$ are difficult to
determine observationally; they probably lie within a factor of two 
of unity (Crutcher 1999), after correcting for likely projection 
effects (Shu et al. 1999). Subcritical clouds with higher $\Gamma_0$ 
contract more slowly, allowing more time for nonaxisymmetric
perturbations to grow. As a result, fragmentation is achieved 
more readily, at a lower surface density. This beneficial effect
of a stronger magnetic field is illustrated in panel f), which
displays the density distribution and velocity field of a cloud 
with $\Gamma_0=2$ (instead of $1.5$) and other parameters identical 
to those of the first model at a time when the maximum surface
density $\Sigma_{\rm max}=10^2$. Clearly, four well-developed
cores have already formed. In contrast, at similar values 
of $\Sigma_{\rm max}$, the cores in the more weakly magnetized 
$\Gamma_0=1.5$ cloud are
either absent (panel c) or barely visible (panel d). Note that 
the positions of the cores are different by about 45 degrees 
in the two models. This is because the more strongly magnetized 
$\Gamma_0=2$ cloud spent 
more time in the quasi-static phase, which enabled an additional
half cycle of oscillation. 

The cores in both models are distributed around a central region of relatively
low surface density and high flux-to-mass ratio, signifying their close
connection to the ring formed in the absence of perturbation (Li 2001). We 
stress
that the ring formation and its fragmentation do not rely on rotation,
which is essential in some non-magnetic models of cloud fragmentation (e.g.
Boss 1996; Klapp \& Sigalotti 1998). The inclusion of a small amount of
rotation, as inferred from molecular line observations of dense cores
(Goodman et al. 1993), enhances the magnetic cloud fragmentation somewhat 
but does not change its behavior fundamentally, as we demonstrate next.
%
%

\subsection{Standard Model with Small Rotation}
\label{rotation}

To gauge the effects of a slow rotation, we repeat the evolution of a
cloud with all parameters identical to those of the first model 
in the last subsection
except the rotation parameter $\omega$, which is set to $0.1$ instead
of zero. Snapshots of the surface density distribution and velocity
field at six selected times are shown in Fig.~2. The general trend of
the cloud evolution is similar to that in the non-rotating case,
with the fragmentation proceeding unimpededly only after the minimum
flux-to-mass ratio drops below the critical value. As before, four
distinct blobs have grown by the end of the computation from the
initially imposed $m=4$ perturbation of the same modest amplitude
$A=0.05$, 
at separations roughly twice those of the non-rotating case. The most
noticeable difference besides the swirling pattern of the velocity
field (i.e., rotation) is that the cores separate out from the 
background at a lower
maximum surface density $\Sigma_{\rm max}$ (compare 
Fig.~1d and Fig.~2e), pointing to
a beneficial effect of the slow rotation on fragmentation. This
effect derives from the fact that rotation retards the radial infall, which
allows more time for the perturbations to grow and fragments to separate
out from the background. We expect the rotation to play an even greater
role on the cloud dynamics as the evolution proceeds to smaller scales.
It should be a crucial ingredient in determining the orbits of the
stars formed. For these reasons, we shall include a slow rotation of
$\omega=0.1$ in all of the models to be discussed hereafter. This
model (especially panel e where $\Sigma_{\rm max}=10^2$), with a set 
of ``standard'' parameters---$r_0=5\pi$, $n=8$,
$\omega=0.1$, $\Gamma_0=1.5$, $A=0.05$ and $m=4$---serves as a
benchmark against which other models will be compared.

\subsection{Different Modes of Perturbation}
\label{order}

Four dense cores are formed in the standard model with an $m=4$ mode of
perturbation. It is natural to ask whether other modes of perturbation
with different values of $m$ can also induce cloud fragmentation and, if 
yes, whether the number of cores produced always matches the value of 
$m$. To address 
these issues, we vary the parameter $m$ and keep all other parameters fixed
at their value in the standard model. The results at a time when the
maximum surface density has increased by two orders of magnitude over
its reference value (i.e., $\Sigma_{\rm max}=10^2$, as in Fig.~2e for
the standard model) are displayed in Fig.~3 for a set of six selected modes
with $m=2$, $3$, $5$, $8$, $11$, and $14$.

The cloud evolution in the $m=2$ case resembles those studied previously
by Nakamura \& Li (2002): soon after the central region becomes
supercritical, the
perturbation grows nonlinearly into a bar \footnote{The true 3D 
structure would be triaxial, as inferred observationally for
dense cores by Jones, Basu \& Dubinski (2001) and Goodwin, 
Ward-Thompson \& Whitworth (2002).} of aspect ratio of roughly 2.
The bar becomes increasingly more elongated as the collapse continues,
reaching an aspect ratio of $\sim 5$ by the time shown in panel a) of
Fig.~3. We have followed the cloud evolution further in time, to a
central density where the isothermal approximation starts to break down.
The bar got more elongated still, but did not break up into pieces.
Therefore, unlike the $m=4$ case, fragmentation into discrete cores does
not occur, at least during the isothermal phase of cloud evolution,
consistent with previous studies of the collapse of non-magnetic filaments
(e.g., Truelove et al. 1997; Sigalotti \& Klapp 2001). In this regard,
the $m=2$ mode is different from the $m=4$ (and other higher order) mode.

The cloud perturbed with an $m=3$ mode appears to evolve in a way
intermediate between the $m=2$ and $m=4$ cases. On the one hand, the
perturbation has become highly nonlinear by the time shown in panel
b), but no discrete blobs are visible, as in the $m=2$ case (panel a).
On the other hand, if we continue the evolution further, the cloud
does eventually break up into (three) discrete cores, as in the
benchmark $m=4$ case (see Fig.~2). We speculate that this intermediate
behavior derives from the fact that the $m=3$ mode distorts more severely
than the $m=4$ mode the ring that would have formed in the absence of any
perturbations (perhaps through preferential central density enhancement) 
but less so than the $m=2$ mode; the breakup into discrete cores 
appears to be intimately linked to ring fragmentation.

Five cores are produced in the $m=5$ case, as shown in panel c). The
core appearances are quiet different from one another, much more so
than for example those in the $m=4$ case, even though the initial
perturbation is periodic azimuthally. We believe that the symmetry
is broken by numerical noises, and amplified by gravity; our
rectangular computational grid appears to have affected the odd modes
(such as the $m=5$ mode shown here and the $m=11$ mode to be discussed
below) more than the even modes (such as the benchmark $m=4$ mode
and the $m=8$ and $14$ modes below). Furthermore, the $m=5$ (and any 
other) mode is not an eigenmode of the (evolving) cloud. Differential 
growth of the eigenmodes contained in the initial perturbation may 
have also contributed to the noted symmetry breaking. 

In each of the three cases with higher order modes ($m=8$, $11$ and $14$),
four blobs are clearly noticeable by the time shown in panels d)-f).
The cores (or rather filaments) are highly elongated, with the potential
of further breakup at higher densities before the isothermal approximation
breaks down. The equal number of fragments in all three cases is interesting,
indicating that the fragment number does not necessarily increase with
the value of $m$. It is limited, in our view, by the number of Jeans masses
contained in the high-density ring in which the filaments are embedded.
Note that the blobs in the $m=14$ case start to separate out from the
background (ring) material at a higher surface density than the other two 
(lower $m$) cases, presumably because the (nonradial gravitational) force
responsible for ring fragmentation is cancelled out to a larger extent
due to more rapid azimuthal variation in the perturbed mass distribution.
The number of fragments depends somewhat on the perturbation amplitude.
For example, eight (rather than four) discrete cores are produced in the 
$m=8$ case at a similar time (when $\Sigma_{\rm max}=10^2$) for a
larger perturbation of fractional amplitude $A=0.3$ (instead of $0.05$).

To summarize, we find that, for the standard cloud, a bar grows out of the
$m=2$ mode, in agreement with our previous finding, and for higher order 
modes discrete cores can be produced. For a modest fractional amplitude 
of $A=0.05$ the number of cores formed ranges from 3 to 5 by the time of
a $10^2$-fold increase in the maximum surface density. The fragmentation
appears to be most effective for an intermediate range of modes. At the 
low$-m$ end (for $m=2$ and to a lesser extent $m=3$), fragmentation
appears difficult, possibly because the surface density increases 
preferentially at the center, which tends to suppress ring formation. 
Fragmentation at the high$-m$ end is limited, on the other hand, by the 
(magnetically modified) Jeans length inside the ring. A larger fractional 
amplitude of the perturbation may change 
this conclusion quantitatively, but not qualitatively. 

We note that the 
elongated multiple dense cores in the $m\ge 3$ cases (especially those 
filament-like ones) as well as the bar in the $m=2$ case will get a 
second chance at fragmentation after the stiffening of equation of state,
producing perhaps an even larger number of discrete fragments that 
could serve as seeds for star formation.

\subsection{Clouds with Different Characteristics}
\label{mass}

So far, we have concentrated on model clouds whose mass distributions
are specified by the pair of parameters $r_0=5\pi$ and $n=8$. We next 
consider different combinations of $r_0$ and $n$. The characteristic 
radius $r_0$ is to be compared with $2\pi$, the (dimensionless) critical 
wavelength for the (thermal) Jeans instability in a disk of uniform 
mass distribution (Larson 1985). The parameter $n$ controls the 
profile of the cloud surface density distribution; those with smaller 
values of $n$ are more centrally peaked.

We first hold the profile parameter fixed at the standard value 
$n=8$, and vary the radius $r_0$. Representative results are shown 
in panels a)-d) of Fig.~4, at a time when the maximum surface density 
$\Sigma_{\rm max}=10^2$, as in Fig.~2e and Fig.~3. Panels a) and b) 
are the snapshots of a cloud with $r_0=10\pi$, twice the standard 
value, perturbed respectively by an $m=4$ and $8$ mode of the 
standard fractional amplitude $A=0.05$. In the $m=4$ case, four dense 
cores have clearly condensed out of a ring-like structure, similar to 
those displayed in Fig.~2e for the standard $r_0=5\pi$ model. The 
main difference is that the cores here have achieved a larger density 
contrast with the background and a higher degree of elongation, 
indicating that the process of fragmentation has proceeded further 
at a similar stage of cloud evolution. The core formation is also more 
dynamic, generating appreciable outward motions which lead to the 
production (well outside the high-density ring) of several secondary 
blobs of local density maxima (not shown). These secondary blobs are 
still magnetically subcritical but appear to be marginally 
self-gravitating. Whether they will eventually collapse onto
themselves and form stars remains uncertain.

In the $m=8$ case (panel b), four discrete over-dense filaments are embedded
in a ring-like structure, similar to the standard $r_0=5\pi$ case (see
Fig.~3d). Unlike those in Fig.~3d, however, the filaments contain two
distinct density peaks (or cores) each, signaling again that the
fragmentation has progressed further. Note that one core in each filament
is much more prominent than the other, even though the initial perturbation
that seeds the core formation is periodic azimuthally. As mentioned earlier,
the asymmetry is probably introduced by numerical noises but amplified
by gravity, especially after the formation of supercritical cores, the
densest of which collapse in a runaway fashion.
In any case, these two examples illustrate that the $r_0=10\pi$ cloud is
more susceptible to fragmentation than the $r_0=5\pi$ cloud. This result
is perhaps to be expected, since a larger characteristic radius
$r_0$ implies a larger number of Jeans masses, which is known to enhance
the fragmentation of non-magnetic clouds. Our calculations demonstrate 
that the same trend holds for strongly magnetized clouds as well.

Reducing the radius $r_0$ is expected to have a negative effect on cloud
fragmentation. This is indeed the case, as illustrated in the next two
panels of Fig.~4, where $r_0=2.5\pi$, half of the standard value. Panel
c) shows that the $m=4$ mode does not grow enough by the time $\Sigma
_{\rm max}=10^2$ that there is no discernible sign of fragmentation. The 
surface density peaks at the cloud center, forming a single dense core, 
just as in the case of no perturbation for this (initially axisymmetric) 
cloud. We have followed the evolution of the cloud perturbed by other 
modes, and found rapid growth only for $m=2$. A snapshot of the $m=2$ 
case is shown in panel d), where the presence of a (twisted) bar is 
evident. It therefore appears that the conditions for the $m=2$ mode 
to grow nonlinearly (into bars) are less stringent than those for the
higher order modes to cause cloud fragmentation. Note that the bar in 
panel d) is less elongated than that of the standard $r_0=5\pi$ cloud 
(Fig.~3a), indicating that a reduction in $r_0$ has hampered not only 
the cloud breakup (for $m\ge 3$) but also bar growth (for $m=2$).

We next consider the influence of the initial surface density profile, as 
specified by the parameter $n$, on cloud fragmentation. For non-magnetic 
clouds, it is well established that those with more peaked density 
profiles are less prone to fragmentation (Boss 1993). We find the same 
trend for strongly magnetized clouds as well. This is illustrated 
in the last two panels of Fig.~4, where $n=2$ (much smaller than the 
standard value of $8$) is chosen, along with the standard radius $r_0=
5\pi$. In the absence of any nonaxisymmetric perturbation, the cloud
collapses into a single central core rather than an off-centered ring, 
as a result of the reduction in $n$, which decreases the number of 
Jeans masses contained in the central plateau region. We find modes 
with $m \ge 3$ do not grow significantly to induce the cloud to break 
up into discrete cores, as exemplified by panel e), where the $m=4$ case 
is displayed. The $m=2$ mode does grow nonlinearly into a bar (panel f), 
although the bar is much less elongated than that in the standard 
$n=8$ cloud (Fig.~3a), whose density profile is much flatter in the 
central region.

To summarize, we find that, as in the non-magnetic case, the ability of
a strongly magnetized cloud to fragment into pieces is controlled mainly
by the number of Jeans masses contained in the central plateau region.
Ring formation under axisymmetry provides a good indicator of whether
a cloud breaks up into fragments or not when perturbed nonaxisymmetrically
with $m > 2$ modes, but not of bar formation from an $m=2$ mode. The 
conditions for bar formation appear to be less stringent than those
for cloud fragmentation. If the characteristic radius $r_0$ and/or profile 
parameter $n$ are small enough so that the central region becomes more 
or less thermally dominated\footnote{In this limit, the thin-disk 
approximation adopted in the paper may break down.}, even bar formation 
will be inhibited (e.g., $r_0=2\pi$ and $n=2$), let alone fragmentation.

\subsection{Random Perturbations and Mode Coupling}
\label{random}

Perturbations in star-forming molecular clouds are not expected to be as
regular as the single modes treated so far, although their exact nature 
is still unknown. In this subsection, we will
consider the opposite extreme, in which the fractional amplitude of
the density perturbation at each grid point is drawn randomly between
$-A$ and $A$. Such random perturbations contain a mixture of modes. The 
number of fragments produced should reflect to some extent the fastest 
growing mode and provide another measure of the ability of a cloud to
fragment.

We begin with a ``canonical'' model in which an $r_0=7\pi$ cloud is
randomly perturbed with a fractional amplitude $A=0.1$. Other 
parameters have their standard values (see Table~1). The cloud 
evolution is shown in the first three panels of Fig.~5, starting with 
the equilibrium state (panel a), where the perturbation is barely 
visible. By the time when the maximum surface density has increased 
by an order of magnitude over its reference value ($\Sigma_{\rm max}
=10$), three condensations of relatively low contrast in surface 
density have formed (panel b). They subsequently fragment into 
five discrete cores, two of which appear to collapse in a runaway 
fashion (panel c). Again, the cores are significantly elongated, 
although not as much as those grown out of single-mode perturbations 
in general, probably as a result of the reduction in the degree of 
symmetry. In this particular model, the $m=5$ mode appears to be the 
fastest growing one among the modes present in the initial random
perturbation. 

We have followed the evolution of randomly perturbed clouds with different 
characteristic radius $r_0$, and found different numbers of fragments. 
The results are 
illustrated in the next two panels of Fig.~5, where $r_0=10\pi$ and $5\pi$ 
are adopted, with everything else fixed as in the canonical model, 
including identical random perturbation at each grid point. Panel d) shows 
that by the time $\Sigma_{\rm max}=10^2$ the same five cores in the 
canonical $r_0=7\pi$ model are also present in the larger/more massive 
$r_0=10\pi$ cloud, 
with similar relative positions. The relative surface densities are 
different, however, with the core at the bottom being the densest (rather 
than the rightmost one). In addition, one core and two condensations
of lower density contrast (and a low-density ``bubble'') are present, 
reinforcing the finding that magnetic clouds with more Jeans masses tend 
to fragment into a larger number of pieces. The smaller/less massive 
$r_0=5\pi$ cloud 
fragmented into three cores (panel e), pointing to the $m=3$ mode as the 
fastest growing one. The result is somewhat surprising, however, since 
the pure $m=3$ mode did not produce any discrete cores by the time 
$\Sigma_{\rm max}=10^2$ for this cloud (see Fig.~3b), whereas other 
modes such as $m=4$ (Fig.~2e) and $5$ (Fig.~3c) did. We suspect that 
nonlinear mode coupling is involved in the formation of these cores, 
perhaps by slowing down the growth of the central surface density that 
may otherwise impede fragmentation. 

To demonstrate mode coupling more clearly, we have repeated the canonical 
model with an additional $m=2$ mode of a small fractional amplitude  
$A=0.01$ superposed. The result is somewhat surprising: by the time 
$\Sigma_{\rm max}=10^2$ (shown in the last panel of Fig.~5), a
double-core has formed, which differs drastically from both the five 
cores formed in the canonical model (panel c) and the 
smooth bar that would grow out of the $m=2$ mode in the absence of 
the random perturbation. It appears that, as in the $r_0=5\pi$ case
shown in panel e), the random perturbation has suppressed the density 
growth near the center, enabling matter to accumulate preferentially 
in two cores rather than a continuous bar. The denser of the cores 
should collapse first into a stellar object (or system). This
should not, however, disrupt the less dense core, since it takes much 
less time for the core material to collapse onto itself than falling 
toward the other center of gravity. As a result, we anticipate that a 
wide binary or hierarchic multiple system (if further fragmentation 
occurs) would be produced, even though the double-core formation is 
not driven by rotation. Higher resolution calculations, perhaps using 
moving sink-cell technique, are needed to ascertain the final 
outcome. We will pursue this intriguing mechanism of (wide) binary 
formation further elsewhere.

To summarize, we have demonstrated that magnetically support clouds of
masses well above the Jeans limit can fragment into discrete dense
cores as a result of ambipolar diffusion, for both single modes and 
random perturbations. The multiple cores are
produced more or less {\it together} (as opposed to in isolation), both
in space and in time. Spatially, they are all located in the magnetically
supercritical region, which is much more compact than the original
cloud. Temporally, they are produced during the supercritical phase
of cloud evolution, which is much shorter than the ``coreless''
subcritical phase. We have shown that more massive magnetic clouds
tend to fragment into a larger number of pieces, as in the nonmagnetic
case. These general features are expected to persist for more realistic
perturbations, which we plan to explore in the future, perhaps with
input from the simulations of turbulent molecular clouds (e.g., 
Ostriker, Gammie \& Stone 1999).

\section{Discussion and Conclusion}

\subsection{Nature of Magnetic Cloud Fragmentation}

The beneficial roles of strong, ordered magnetic fields in resolving the
thorny
``angular momentum problem'' of star formation and in preventing the
overall dynamical collapse of molecular clouds are well known (Shu et al.
1987; Mouschovias \& Ciolek 1999). Their effects on cloud fragmentation 
are less explored, and appear
more subtle. On the one hand, the magnetic pressure can assist the thermal
pressure in erasing density inhomogeneities and thus impede
fragmentation. On the other hand, the magnetic tension force dilutes
gravity, allowing more time for perturbations to grow and over-dense
fragments to separate out (see also Boss 2002). The
deleterious effects are brought out clearly in the simulations of
Phillips (1986a), where the (frozen-in) magnetic fields are taken to
be uniform initially, with zero tension force. Phillips (1986a)
followed the collapse of a set of such uniformly magnetized clouds
with a range in the ratio of
thermal to gravitational energies and degree of magnetization. He
found that none of the clouds fragmented, even in the presence of a
large, $50\%$ density perturbation. We believe that the lack of
fragmentation is mainly due to the weakness of the magnetic tension
force relative to gravity in his models, which does not slow down
the nearly free-fall collapse appreciably. 
The tension force is stronger in the models of Phillips (1986b)
where the initial fields are non-uniform. However, nonaxisymmetric
perturbations are not imposed on these clouds, and it is not clear
whether fragmentation is enhanced by the stronger tension force or
not. The beneficial effects of magnetic tension force on fragmentation
is illustrated most cleanly with the thin-sheet model of Shu \& Li
(1997), which is threaded everywhere by a (frozen-in) magnetic field
of critical strength. In such an idealized cloud, the gravity is
exactly balanced by the tension force for arbitrary mass distributions,
including those with large numbers of condensations (see Allen \& Shu
2000 for a toy model of the fragmentation of critically magnetized
clouds induced by protostellar winds).

The opposing effects of magnetic fields on cloud fragmentation call for
a numerical attack, especially in the presence of ambipolar diffusion,
which changes the relative importance of the two with time. Our
calculations demonstrate that, despite these extra complications, the
ambipolar diffusion-driven fragmentation of magnetically subcritical
clouds has much in common with that of non-magnetic clouds (e.g.,
Boss 1996; Klapp \& Sigalotti 1998): those clouds containing more
Jeans masses and
having flatter mass distributions are more susceptible to fragmentation.
These trends can be understood qualitatively from the classical Jeans
analysis (e.g., Larson 1985) and its extension to lightly ionized,
strongly magnetized media (Langer 1978). In both cases, cloud
fragmentation involves multiple Jeans masses and is driven by the
(magnetically diluted) gravity over the resistance of the (magnetically
enhanced) pressure gradient. A major advantage of magnetic clouds over 
their non-magnetic counterparts is their ability to contain multiple
Jeans masses without collapsing promptly. Equally important to 
fragmentation is the flattening of mass distribution along field
lines, which enables self-gravitating off-center pockets to 
preferentially collapse onto themselves rather than falling toward 
the center (Larson 1985). 

Once the central region of a magnetically supported, multi-Jeans mass
cloud has become supercritical, nonaxisymmetric perturbations start to
grow. The time for growth is lengthened considerably by the magnetic
tension force, which remains strong enough during the initial period of 
the supercritical phase to balance out the gravity to a large extent.
Nakamura \& Li (2002) demonstrated that m=2 density perturbations of
modest fractional amplitude of $5\%$ can grow nonlinearly into bars 
of aspect ratio $\sim 2$ during this period (see \S~\ref{results} for 
more examples). It is also during this period that cloud breakup into 
multiple cores occurs, as speculated by Li (2001) and shown explicitly 
in the paper. This critical phase of cloud evolution enables the 
(initially) magnetically subcritical clouds to either break up directly 
into pieces and/or become significantly elongated, setting the stage 
for possible (further) fragmentation at later times. It is absent from 
the evolution of both the non-magnetic and (strongly) magnetically 
supercritical clouds. The beneficial effects of a strong magnetic field 
on fragmentation are illustrated vividly in Fig.~1, which shows that 
increasing the field strength of a subcritical cloud makes it easier
for the cloud to break up into multiple cores. 

Fragmentation of strongly magnetized cloud cores has been examined by 
Boss in a series of papers (see Boss 2002 for the latest). Using a 3D 
code that treats the magnetic forces and ambipolar diffusion 
approximately, he finds that prolate magnetic cores tend to produce 
binaries and oblate cores multiple systems. Our calculations are 
similar to his oblate core calculations but with a different focus: 
the formation of multiple dense cores out 
of a flattened, more diffuse background cloud
\footnote{Even though our discussions of the dimensionless solutions 
of cloud evolution are centered on relatively low initial densities 
appropriate for pre-core conditions, it may be possible to reinterpret 
these solutions using (oblate) dense cores as the starting point (if 
such cores are magnetically subcritical). In this reinterpretation, 
the discrete overdense blobs developed toward the end of our 
calculations would have much higher densities and smaller separations, 
comparable to the multiple protostellar fragments found by Boss (2002). 
The fact that these two very different sets of calculations (different 
geometries, numerical methods and treatments of radiative transfer, 
magnetic field and ambipolar diffusion) yield qualitatively similar 
results regarding fragmentation is reassuring.}. We concur with Boss's 
general conclusion that fragmentation can take place {\it despite} the 
presence of significant magnetic fields. Indeed, one may go one step 
further, and speculate that the fragmentation of relatively quiescent 
clouds into binaries, multiple systems and small groups occurs largely 
{\it because of} the strong magnetic field, although more detailed 
calculations are needed to back up this viewpoint.

\subsection{Implications on Small Stellar Group Formation}

Magnetically supported clouds can have masses well above the Jeans limit
and a naturally-flattened mass distribution, both of which are conducive 
to the growth 
of nonaxisymmetric perturbations as the magnetic support weakens with 
time due to ambipolar diffusion. Nonlinear growth of 
perturbations produces two basic types of over-dense structure in the 
supercritical phase of cloud evolution: single, isolated bars for the 
$m=2$ mode and, for higher order modes, discrete multiple cores, each 
of which is also significantly elongated in general. Nakamura \& Li (2002)
have speculated that the elongated dense cores (or bars), with their 
elongation strongly amplified at higher densities by the Lin-Mestel-Shu 
(1965) instability, may fragment during the subsequent adiabatic phase 
of cloud evolution, individually producing perhaps a binary or
multiple stellar system. More detailed calculations are needed to put
the speculation on a firmer ground. Here, we wish to concentrate on the
implications of the multiple cores, whose collapse to form collectively
a small group of stars appears more certain. This is because by the end
of the computations (some of) the cores have cleanly separated out from
the background and from one another, and are well on their way to produce
stellar density objects.

Some of the salient features of multiple core formation can be seen from
Fig.~2, where the evolution and fragmentation of the benchmark cloud with
a slow rotation are displayed. Adopting a fiducial temperature of $10$~K and
visual extinction of $A_V=1$, we find that the cloud initially contains a
central region of flat surface density distribution of radius $\sim 1$~pc
and mass $\sim 50$~$M_\odot$. It takes about 5.2 Myrs for the cloud to reach
the supercritical state, and another 2.4 Myrs for the m=4 perturbations
to grow into well-defined multiple dense cores. The cores are separated
by $\sim 0.05$~pc and, at the end of the calculation, have peak extinction
of over 300. They are embedded within a much larger common envelope of
lower visual extinction. The cores are well on their way to dynamic
collapse from inside-out, each expected to form a protostar or protostellar
system which grows by accreting mass competitively from the envelope. In
this particular example, we anticipate the formation of a group of stars
within a relatively small region of $0.05$~pc or less, although detailed
properties of the group, such as the final masses of individual members
and their orbital elements, are unknown. Our calculation sets the stage
for future higher resolution calculations capable of following the
subsequent core evolution to the formation of point mass (protostar) and
beyond, which would help elucidate such important topics as competitive
mass accretion and orbital evolution of the (still accreting) protostars
(Bonnell et al. 1997). Substantial orbital evolution is expected of the
group members, not only because the dense cores---the raw materials out
of which they are formed---are not rotationally supported (from collapsing
into one another), but also because the stars can interact gravitationally
with the envelope (by raising tides) and with one another. The former
tends to shrink the stellar orbits, and the latter may lead to the
ejection of some (lighter) members while leaving others more tightly bound.
The decay of small-number groups may be an important channel for producing
binaries and multiple systems (e.g., Sterzik \& Durisen 1999) and, in
the scenario of Reipurth \& Clarke (2001), be responsible for brown dwarf
formation.

How are the magnetically subcritical clouds (or ``clumps'' following the
notations of Williams et al. 2000) of multiple Jeans masses capable of
fragmentation created in the first place? We do not have a definitive
answer, but suspect that it probably involves in one way or another
turbulence, which is strong in the low density regions out of which the
subcritical clouds or clumps are condensed. The condensation could occur
for example through compression in the converging regions of a turbulent
flow (e.g., Ostriker et al. 1999), and/or as a result of localized 
turbulence decay (Myers 1999). Since the turbulence
is highly supersonic, the thermal pressure (and thus the Jeans limit)
should have little relevance to the process. One therefore expects clouds
or clumps of more than one Jeans mass to be a common product, at least 
for those more massive, self-gravitating ones that are the sites of
star formation. Whether
such clouds or clumps are magnetized strongly enough to be subcritical
or not is unclear. As yet, there is not enough direct Zeeman measurements
of field strength in relatively low density regions to provide a firm
answer to this crucial question one way or the other (Crutcher 1999).
In the the standard scenario of isolated low-mass star formation the
clumps are envisioned to be magnetically subcritical (Shu et al. 1987;
Mouschovias \& Ciolek 1999). If this turns out to be the case, then their
propensity for containing more than one Jeans mass would make them 
suitable for group formation.

There is some observational evidence that small groups are a common product
of low-mass star formation, in both the so-called ``isolated'' and
``clustered'' modes (Shu et al. 1987). The archetype of the former is
the Taurus molecular cloud. Molecular line surveys of this nearby dark
cloud in $^{13}$CO (Mizuno et al. 1995; at a spatial resolution of
$\sim 0.1$~pc), C$^{18}$O (Onishi et al. 1996; $\sim 0.1$~pc) and
H$^{13}$CO$^+$ (Onishi et al. 1999; $\sim 0.01$~pc) have found an
apparent threshold in the H$_2$ surface density of $\sim 8\times
10^{21}$~cm$^{-2}$ for ongoing star formation, as marked by either
dense compact (pre-stellar) H$^{13}$CO$^+$ cores or ``cold'' (in IR
colors) young stellar objects (YSOs; Onishi et al. 1998). Interestingly,
those star-forming C$^{18}$O cores above the threshold typically
contain more than one compact core and/or cold YSO, with the more
massive ones associated with a larger number of cold objects. Higher
resolution continuum maps have also revealed many examples of dust
condensations in close proximity, with separations of order
$\sim 0.1$~pc or less (e.g., Looney, Mundy \& Welch 2000; Shirley
et al. 2000; Motte \& Andre 2001). These groupings of pre-stellar
objects and YSOs in relatively quiescent regions could result 
from the ambipolar diffusion-driven
fragmentation of magnetically supported clouds discussed in this
paper. Even more striking examples of such groupings can be found in
cluster-forming regions, such as the fragmented ring-like structure
in the Serpens cloud core (Williams \& Myers 2000) and the dozen or
so starless fragments in the $\rho$ Oph B2 core (Motte et al. 1998).
Whether they could be produced in the same manner is less certain,
complicated by the strong turbulence, which is present in these regions
but not treated explicitly in our calculations.

One prediction of our scenario is that the group members would have rather 
small spread in age even though they are formed through ambipolar diffusion, 
a process that 
can take up to 10 Myrs. The reason is that the dense cores are produced
in our picture together during the supercritical phase, which lasts for 
a small fraction of the total time of cloud evolution. Even though the
denser of the cores would collapse to form stars first, the spread in 
stellar age
should still be relatively small, perhaps comparable to that of stars 
formed from the fragmentation of non-magnetic clouds---much less than the
spread of those formed from separate subcritical clouds in
complete isolation from one another.

\subsection{Conclusion and Future Work}

To summarize, we have followed numerically the ambipolar diffusion-driven
evolution of magnetically supported molecular clouds in the presence
of various nonaxisymmetric modes of perturbations. We confirmed our
previous finding that the $m=2$ perturbations of modest amplitude grow
readily into bars, which have implications for binary and multiple
system formation. Clouds perturbed with higher order modes can break up
into multiple cores, which should collapse to form collectively small
groups of stars. These calculations are the first step toward a 
comprehensive theory of multiple star formation in strongly magnetized
clouds. 

Our calculations are limited to the isothermal regime. They will be
extended into the adiabatic regime at higher densities, to follow the
expected breakup of the (long) bars and multiple cores, which are also 
significantly elongated in general. A new feature at the higher densities
is the magnetic decoupling (e.g., Nishi et al. 1991), which will
require a more elaborate treatment of the magnetic coupling than the
one we adopted. Ultimately, we would like to continue our calculations
into the protostellar accretion phase, hoping to determine the masses
of individual stars and their orbits, perhaps with an approximate
treatment of the magnetic braking which may significantly affect the
angular momentum evolution of the system.

Numerical computations in this work were carried out 
 at the Yukawa Institute Computer Facilities, Kyoto University.
F.N. gratefully acknowledges the support of the
JSPS Postdoctoral Fellowships for Research Abroad. We thank the
referee for a prompt and helpful report.


\begin{deluxetable}{lllllll}
\tablecolumns{7}
\tablecaption{Parameters of Models \label{table1}}
\tablewidth{5.5in}
\tablehead{
  \colhead{$r_0$}     & \colhead{$n$}          & \colhead{$\omega$}
 &\colhead{$\Gamma_0$}  & \colhead{perturbation} & \colhead{$A$}
 &\colhead{display}
}
\startdata
$5\pi$   & 8 & 0.0 & 1.5 & $m=4$  & 0.05 & Fig.1a-e\\
$5\pi$   & 8 & 0.0 & 2.0 & $m=4$  & 0.05 & Fig.1f\\
$5\pi$   & 8 & 0.1 & 1.5 & $m=4$  & 0.05 & Fig.2a-f~$^*$ \\
$5\pi$ & 8 & 0.1 & 1.5 & $m=2, 3, 5, 8, 11, 14$ & 0.05 & Fig.3a,b,c,d,e,f\\
$10\pi$  & 8 & 0.1 & 1.5 & $m=4, 8$ & 0.05 & Fig.4a,b \\
$2.5\pi$ & 8 & 0.1 & 1.5 & $m=4, 2$ & 0.05 & Fig.4c,d \\
$5\pi$   & 2 & 0.1 & 1.5 & $m=4, 2$ & 0.05 & Fig.4e,f \\
$7\pi$   & 8 & 0.1 & 1.5 & random & 0.1 & Fig.5a-c~$^{**}$ \\
$10\pi$   & 8 & 0.1 & 1.5 & random & 0.1 & Fig.5d \\
$5\pi$   & 8 & 0.1 & 1.5 & random & 0.1 & Fig.5e \\
$7\pi$   & 8 & 0.1 & 1.5 & random and $m=2$ & 0.1 and 0.01 & Fig.5f
\enddata
\tablecomments{All models are computed on an initial grid of $256\times 
256$ in the disk ($x$-$y$) plane. The grid number is increased to 
$512\times 512$ for each level of refinement. The third model from 
the top (marked by $^*$) is the ``standard'' model against which 
others are compared, and the ``canonical'' randomly-perturbed model 
is marked by $^{**}$. 
}
\end{deluxetable}

\clearpage
\begin{figure}
\epsscale{0.90}
\plotone{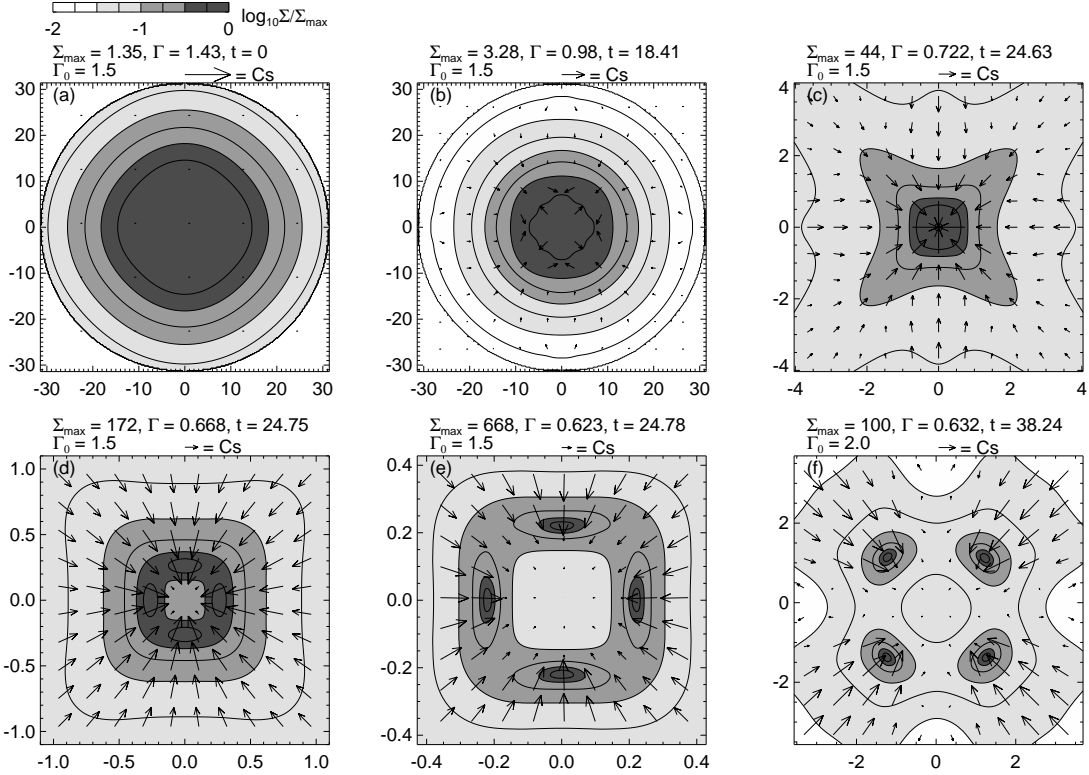}
\caption{ Snapshots of the 
surface density distribution and velocity field for two strongly magnetized,
non-rotating clouds  perturbed by an $m=4$ mode of fractional amplitude 
$A=0.05$. 
Panels (a) through (e) show the evolution and fragmentation of a cloud 
specified by parameters $(r_0, n, \omega, \Gamma _0)=
(5\pi, 8, 0, 1.5)$, and panel (f) illustrates the beneficial effect
of a stronger magnetic field (with $\Gamma _0 = 2.0$ rather than 
$1.5$) on fragmentation. 
In panels (c) through (f), only the central regions are shown.
The contours in each panel are for the surface density normalized by 
$\Sigma_{\rm max}$, whose value is given above each panel. Also
given are the flux-to-mass ratio ($\Gamma$) at the
density peak, dimensionless evolution time ($t$), and
initial flux-to-mass ratio ($\Gamma_0$).  The arrows are velocity
 vectors normalized by the effective isothermal sound speed $c_s$ 
(without magnetic contribution), whose magnitude is indicated 
above each panel. Note that in panel (a) the added perturbation 
lowers the flux-to-mass ratio $\Gamma=1.43$ from the reference value 
$\Gamma_0=1.5$ by $5\%$, corresponding to the fractional amplitude 
of the perturbation. For dimensional units, see subsection \S~\ref{methods}.
\label{fig:1}}
\end{figure}

\begin{figure}
\plotone{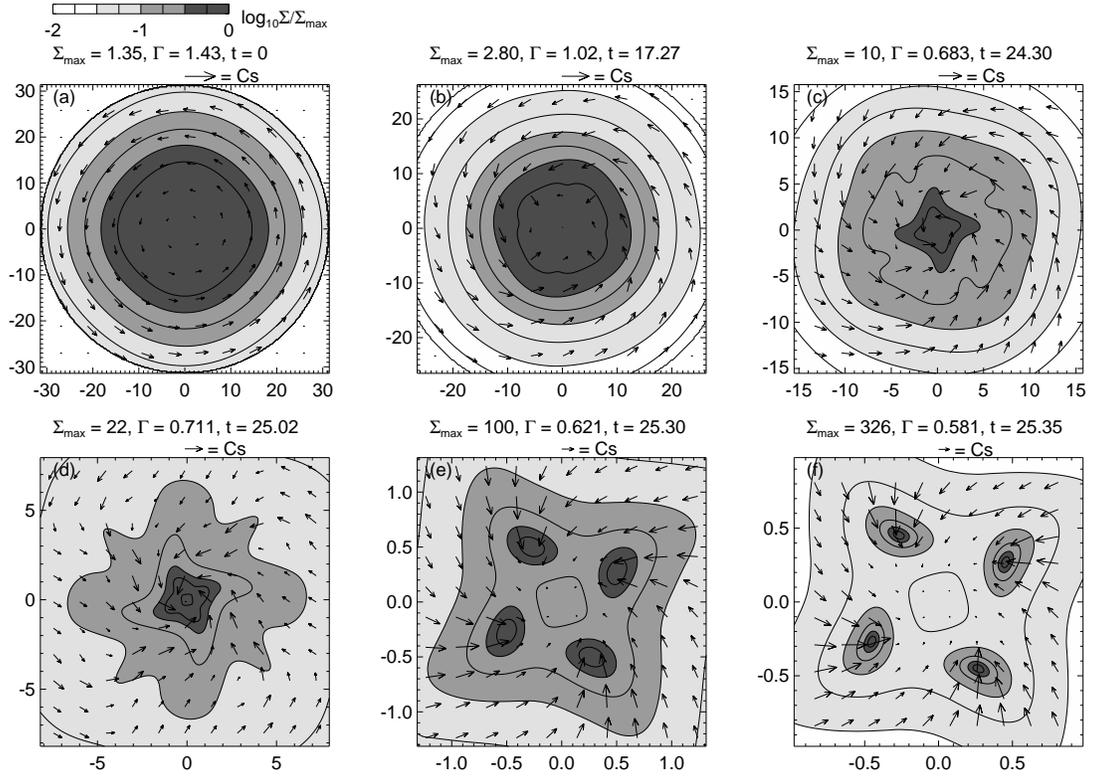}
\caption{Snapshots of the surface density distribution and velocity
field of the ``standard'' model with parameters $(r_0, n, \omega, 
\Gamma _0, A)=(5\pi, 8, 0.1, 1.5, 0.05)$, showing the growth of an $m=4$ 
perturbation in a slowly-rotating cloud leading to fragmentation. 
The contours, arrows and notations have the same meaning as in 
Fig.~\ref{fig:1}.
\label{fig:2}}
\end{figure}

\begin{figure}
\plotone{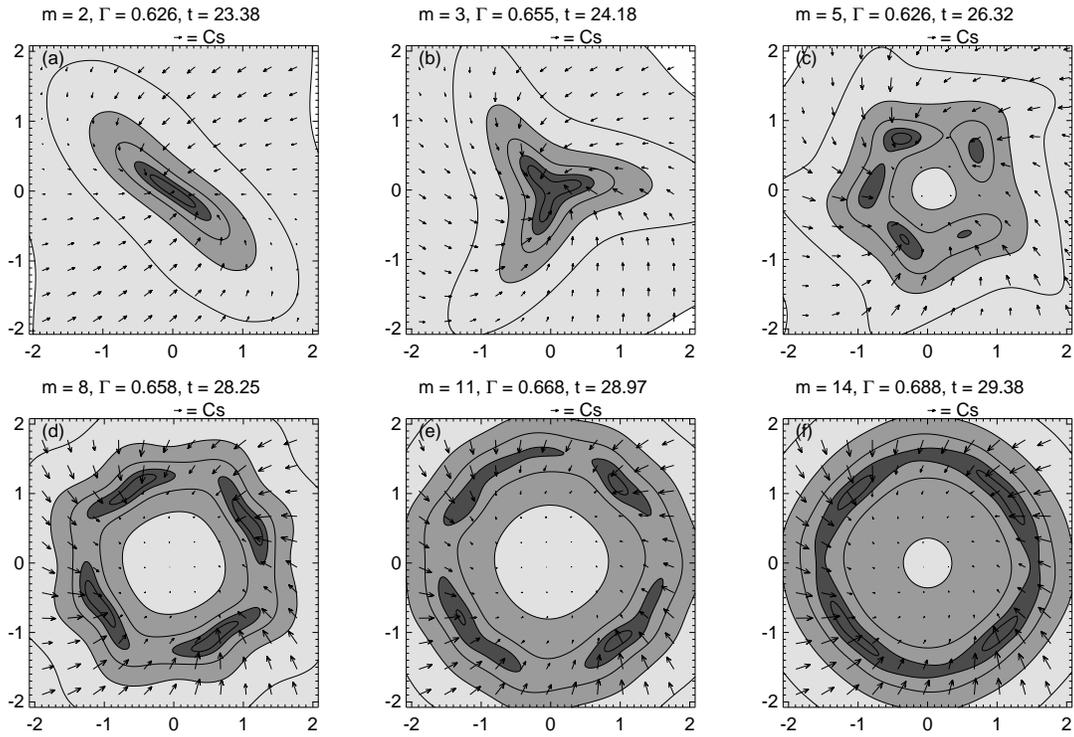}
\caption{Snapshots of the surface density distribution and velocity
field at a time corresponding to $\Sigma_{\rm max}=10^2$ for the 
``standard'' cloud perturbed by six selected modes of standard 
fractional amplitude $A=0.05$. The imposed modes are $m=2$ (a), 3 
(b), 5 (c), 8 (d), 11 (e), and 14 (f). The contours and arrows 
have the same meaning as in Fig. \ref{fig:1}, and the numbers above 
each panel denote the mode of perturbation ($m$), flux-to-mass ratio 
($\Gamma$) at the density peak, and dimensionless evolution time 
($t$). The connection between fragmentation and ring formation is 
evident.
\label{fig:3}}
\end{figure}

\begin{figure}
\plotone{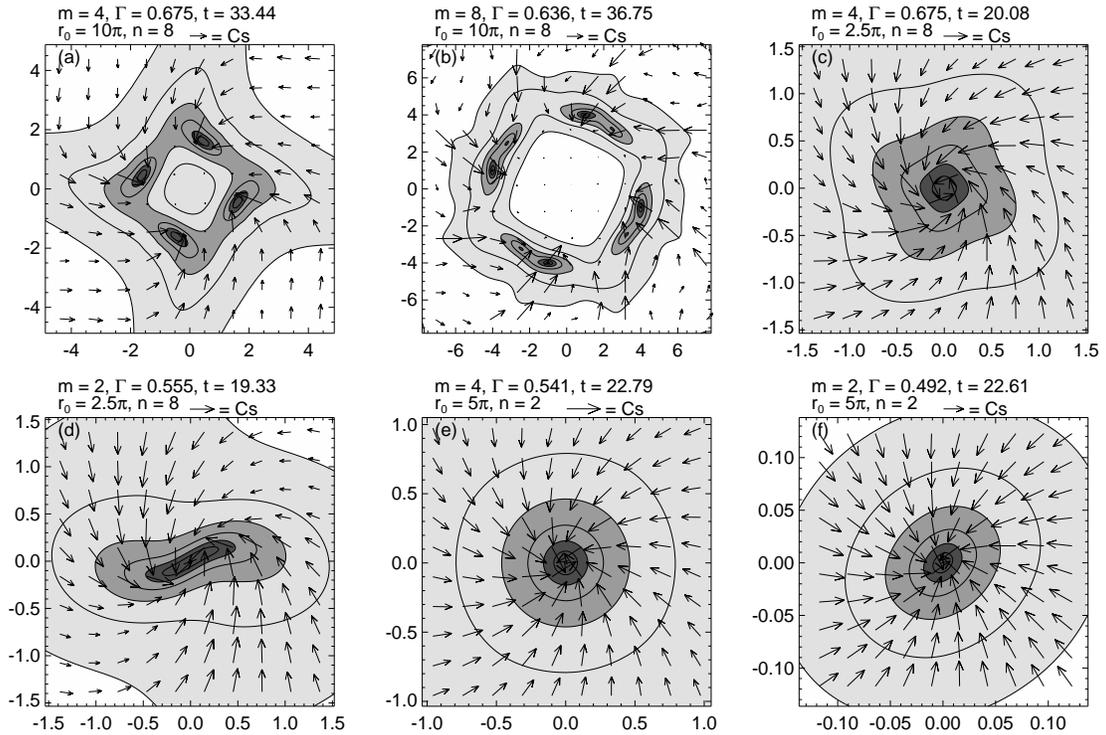}
\caption{Snapshots of the evolution of three clouds at a time when 
$\Sigma_{\rm 
max} = 10^2$, showing the beneficial effects of a larger characteristic
radius $r_0$ on cloud breakup (panels a and b), and the negative effects 
of a smaller $r_0$ (panels c and d) or $n$ (more centrally peaked 
density distribution; panels e and f) on fragmentation and bar formation.
The contours and arrows are the same as in Fig. \ref{fig:1}. The models
are specified by the numbers above each panel (see also Table~1). 
\label{fig:4}}
\end{figure}

\begin{figure}
\plotone{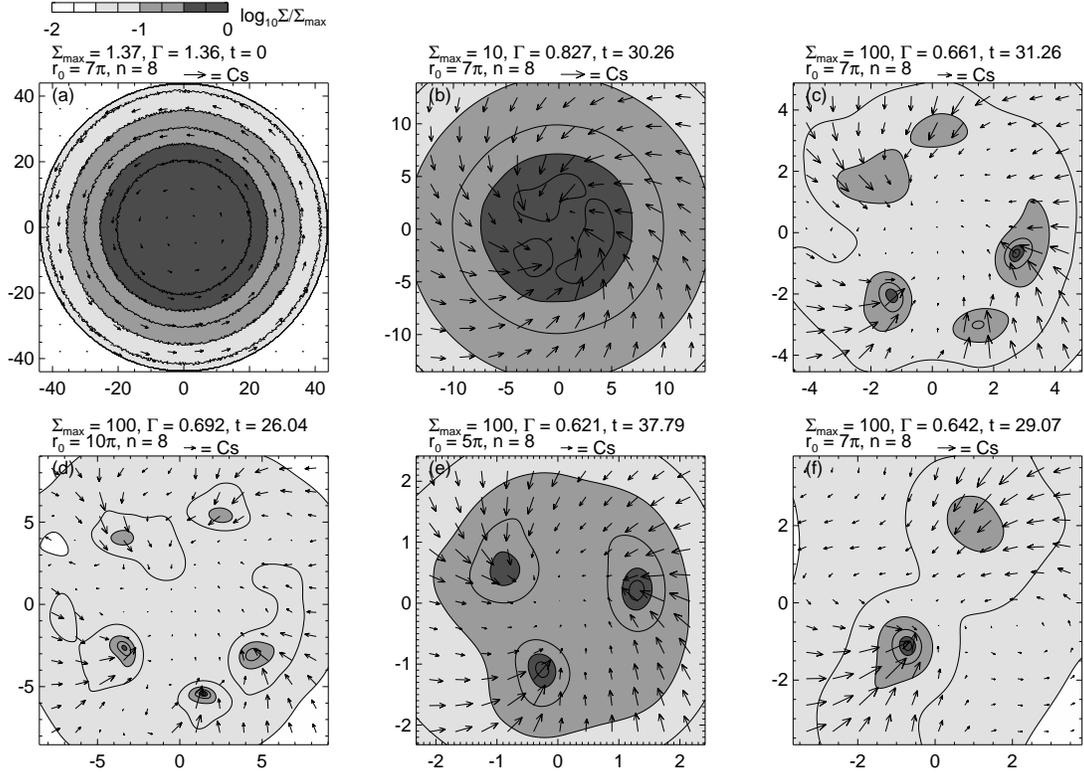}
\caption{Snapshots showing the evolution and fragmentation of an $r_0=7\pi$ 
cloud in the presence of a random density perturbation of fractional 
amplitude $A=0.1$ (panels a-c; the ``canonical'' model), and the effects 
of varying characteristic radius ($r_0=10\pi$ and $5\pi$ for panel d and 
e respectively) on the number of fragments obtained. In panel f), an $m=2$ 
mode of very small fractional amplitude $A=0.01$ is superposed on the random 
perturbation of the canonical model, leading to the production of a 
double-core, which 
may have implications for wide binary formation. The contours, arrows and 
notations have the same meaning as in Fig. \ref{fig:1} and other figures.
\label{fig:5}}
\end{figure}
\end{document}